\documentstyle[epsfig,12pt]{article}

\makeatletter
\let\chapter\hid@chapter
\makeatother

\def\L{{\cal L}}
\def\sframe{\hbox{$\underline{S\ }\!\!|$\ }}
\def\eframe{\hbox{$\underline{E\ }\!\!|$\ }}
\def\half{\hbox{\small $\frac{1}{2}$}}
\def\NPB{{\em Nucl. Phys.} B}
\def\PLB{{\em Phys. Lett.}  B}

\begin{document}
\rightline{\small BGU-PH-97/13}

\ 

\centerline{\Large Graceful Exit in String Cosmology }

\ 

\centerline{\large Ram\,Brustein${}^{(1)}$ and Richard\,Madden${}^{(2)}$ }

\ 

\centerline{\small \it
Department of Physics, Ben-Gurion University, 84105 Beer-Sheva, Israel}

\centerline{\small \tt  e-mail: ramyb, madden @bgumail.bgu.ac.il}


\

\

\

\noindent
\begin{abstract}
The graceful exit transition from a dilaton-driven inflationary phase
to a decelerated Friedmann$-$Robertson$-$Walker era requires certain
classical and quantum corrections to the string effective action.
Classical corrections can stabilize a high curvature string 
phase while the evolution is still in the weakly coupled regime, 
and quantum corrections can induce violation 
of the null energy condition, allowing  
evolution towards a decelerated 
phase.  
\end{abstract}
\vspace{2.5in}
\hrule 

\ 

\noindent
1. Based on talk given at the workshop Modern Modified Theories of 
Gravitation and Cosmology, 29-30 June 1997, Ben-Gurion University, 
Beer-Sheva, Israel.\\
2. Based on talk given at the International Europhysics Conference on
High Energy Physics, 19-26 August 1997, Jerusalem, Israel.

\newpage
\section{Introduction}

String theory predicts gravitation, but the 
gravitation it predicts is not that of standard general relativity. 
In addition to the metric fields, string gravity also contains a scalar dilaton,
that controls the strength of coupling parameters. 
An inflationary scenario \cite{gv1}, is based on the fact that
cosmological solutions to string dilaton-gravity 
come in duality-related pairs, an inflationary branch in which the Hubble 
parameter increases with
time and a decelerated branch that can be connected smoothly to a standard
Friedmann$-$Robertson$-$Walker (FRW) expansion of the Universe with
constant dilaton. The scenario (the so called ``pre-big-bang") is that
evolution of the Universe starts from a state of very small curvature and
coupling and then undergoes a long phase of dilaton-driven kinetic inflation
and at some later time joins smoothly standard 
radiation dominated cosmological evolution, thus giving rise to a singularity
free inflationary cosmology. 

However, in the lowest order effective action these two branches 
are separated by a singularity. Additional fields or correction terms 
need to be added to make this ``graceful exit'' transition possible.
It has been studied intensely in the last few years, but all
attempts failed to induce graceful exit. 
In \cite{bv, kmo} it was shown that the transition is forbidden  
for a large class of fields and potentials.
In \cite{BM} we proposed to use an effective description in terms of sources
that represent arbitrary corrections to the lowest order equations 
and were able to formulate a set of necessary 
conditions for graceful exit and to relate them to energy conditions
appearing in singularity theorems of Einstein's general relativity \cite{he}. 
In particular, we showed that a successful exit requires violations of the
null energy condition (NEC) and that this violation is associated with
the change from a contracting to an expanding universe (bounce) in the 
``Einstein frame'', defined by a conformal change of variables.
Since most classical sources obey NEC this
conclusion hints that quantum effects, 
known to violate NEC in some cases, may be
the correct sources to look at.  

Because the Universe evolves towards higher curvatures and
stronger coupling, there will be some time when the lowest order effective
action can no longer reliably describe the dynamics and it must be corrected.
Corrections to the lowest order effective action come from two sources. The
first are classical corrections, due to the finite size of strings, arising
when the fields are varying over the string length scale
$\lambda_s=\sqrt{\alpha'}$. These terms are important in the regime of large
curvature.  The second are quantum loop corrections. The loop expansion is
parameterized by powers of the string coupling parameter $e^\phi=g_{string}^2$,
which in the models that we consider is time dependent.  So
quantum corrections will become important when the dilaton $\phi$ 
becomes large, the regime we refer to as strong coupling.  

In \cite{BM2} we were able to find an explicit model that satisfies 
all the necessary condition and to produce the first example of a 
complete exit transition. The specific model we present here makes 
use of both classical and quantum corrections. 
We allowed ourselves the freedom to choose the
coefficients of correction terms which generically appear in string effective
actions. Our reasoning for allowing this
stems in part from a lack of any real string calculations and in part
by our desire to verify, by constructing explicit examples, 
the general arguments of \cite{BM}. 

\section {A Specific Exit Model}
 
String theory effective action in four dimensions 
takes the following form in the string frame (\sframe), 
\begin{equation} 
S_{eff}^{{\sframe}}= 
\int d^4 x \left\{ \sqrt{-g}\left[ \frac{e^{-\phi}}{16 \pi \alpha'} 
\left(R+\partial_\mu\phi \partial^\mu\phi\right)\right]+ 
\half \sqrt{-g} \L_c\right\}, 
\label{effacts} 
\end{equation} 
where $g_{\mu\nu}$ is the 4-d metric and $\phi$ is the dilaton.  
The Einstein frame (\eframe) is defined by the change of variables
$g \rightarrow e^{\phi/2} g$, diagonalizing the metric and dilaton 
kinetic terms and resulting in equations of motion similar to those of 
standard general relativity. 

We are interested in solutions to the equations of motion derived from  
the action (\ref{effacts}) of the FRW type with vanishing spatial curvature 
$ds^2= -dt_S^2+a_S^2(t) dx_i dx^i$ and $\phi=\phi(t)$.  
The contribution of $\L_c$ is contained in 
the correction energy-momentum  
tensor $T_{\mu\nu}=\frac{1}{\sqrt{-g}} 
\frac{\delta \sqrt{-g} \L_c}{\delta g^{\mu\nu}}$,  
which will have the form $T^\mu_{\ \nu}=diag(\rho,-p,-p,-p)$. In addition 
we have another  source term arising from the  $\phi$  
equation, $\Delta_\phi\L_c=\half \frac{1}{\sqrt{-g}}   
\frac{\delta \sqrt{-g} \L_c}{\delta\phi}$.

The $00$ equation of motion is quadratic and may be conveniently
written, 
\begin{equation} 
\dot\phi=3 H_S \pm \sqrt{3 H_S^2+e^{\phi} \rho},
\label{rhoeq}
\end{equation} 
where $H_S=\dot a_S /a_S$, and we have fixed  our 
units such that $16\pi\alpha'=1$. The choice of sign here corresponds
to our designation of (+) and ($-$) branches.  
For the corrections we propose classical, one and two loop terms of 
a particularly simple form. The first term in (\ref{correction}) is 
the form of $\alpha'$ corrections
examined in \cite{gmv}, the second and third are plausible forms for the
one and two loops corrections respectively. The large coefficients 
account for the expected large number of degrees of freedom contributing
to the loop. The signs of these terms are deliberately chosen to force
the exit. 
\begin{equation} 
\half \L_{c}=e^{-\phi}
                   (\frac{R_{GB}^2}{4}- \frac{(\nabla \phi)^4}{4}) -
             1000 (\nabla \phi)^4 + 1000 e^{\phi} (\nabla \phi)^4.
\label{correction}
\end{equation} 
We have checked that qualitatively similar evolution is obtained for a 
range of coefficients, of which (\ref{correction}) is a representative.

We set up initial conditions in
weak coupling near the (+) branch vacuum and a numerical integration yields
the evolution shown in Fig. 1  in the $\dot \phi$,$H_S$ phase space.
We have also plotted lines marking important landmarks in the evolution,
the (+) and ($-$) vacuum ($\rho=0$ in (\ref{rhoeq})), the line of branch 
change $(+) \rightarrow (-)$ (square root vanishing in (\ref{rhoeq})) 
and the position
of the \eframe bounce ($H_E=0$, $\dot \phi=2 H_S$). We see the that the
evolution falls into three distinct phases which we will discuss in 
turn.
\begin{figure}[h]
\begin{center}
\epsfig{file=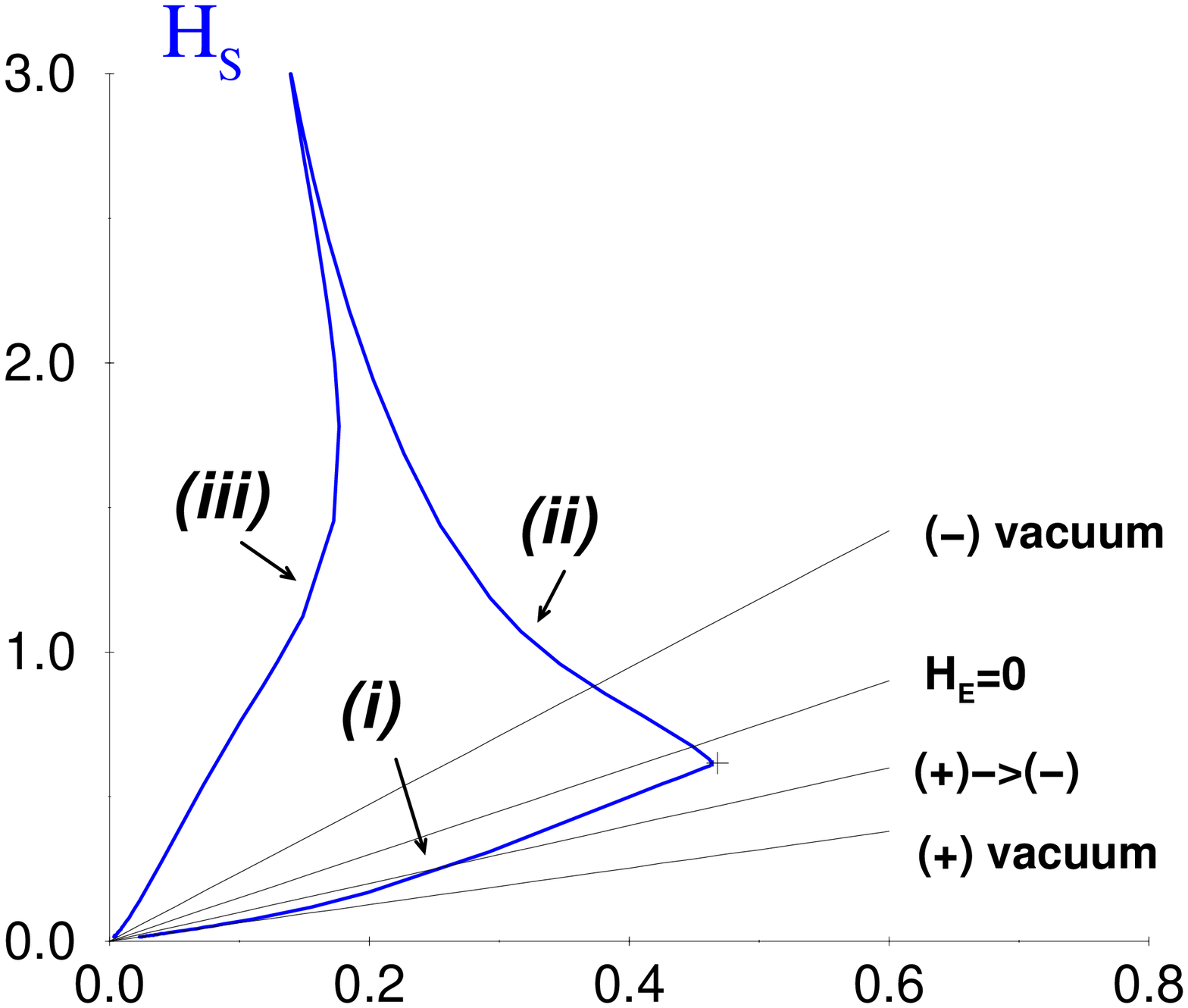,width=10cm,height=7cm,bbllx=12,bblly=-20,
        bburx=576,bbury=480}
\end{center} 
\caption{The correction induced graceful exit}
\end{figure}
{\bf Phase (i)}
The solution begins with a long evolution near the (+) vacuum, this
is the inflationary phase. As curvature becomes large we see deviation
induced by the $\alpha'$ corrections in (\ref{correction}) and without
influence from other corrections the solution would settle into the fixed 
point noted in \cite{gmv}, marked with a '+'. 
The solution does cross 
the line of branch change, but does not execute the \eframe bounce 
required by a complete exit, corresponding to the fact it does not 
violate NEC in the \eframe.
{\bf Phase (ii)}
While the Universe sits near the fixed point the dilaton is still increasing 
linearly, so eventually the loop corrections in (\ref{correction}) 
will become important. The first to do so is the one loop correction.
Since we require further NEC violation to complete the \eframe bounce,
we have chosen the sign of the one loop correction to provide this 
and in this phase corrections are dominated by this term. As a result a 
 bounce occurs and the evolution proceeds into the $\rho>0$ region. 
We checked that other forms of loop correction will have the same effect
if they are introduced with a coefficient allowing NEC violation. But without
further corrections this solution would continue to grow into regions
of larger curvature and stronger coupling. We refer to this era as
``correction dominated'' and we also find there are obstacles to 
stabilizing the dilaton with standard mechanisms like capture in a 
potential or radiation production.
{\bf Phase (iii)}
To offset the destabilizing NEC violation we have introduced the 
two loop correction with the opposite sign, allowing it to overturn
the NEC violation when it becomes dominant as $\phi$ continues to grow.
Indeed during this phase we see the expansion decelerating, dilaton 
growth stabilizing, and the corrections vanishing.
We have also checked that in this phase the dilaton can be 
captured into a potential minimum or halted by radiation production.
This phase can be smoothing joined to standard cosmologies.

\section {Conclusion}

Corrections to lowest order string effective action
can induce graceful exit.
It remains to be seen whether
string theory  has the predictive power to fix the form and coefficients of 
these corrections and thus give a definite answer to the question of whether 
graceful exit indeed proceeds as suggested in our model.

\section{Acknowledgment}
Research supported in part by the Israel Science Foundation administered by the Israel 
Academy of Sciences and Humanities.

%


\begin{thebibliography}{99}
\bibitem{gv1} G. Veneziano, \PLB 265  (1991) 287; 
M. Gasperini and G. Veneziano, {\em Astropart. Phys.} 1 (1993)  317.
\bibitem{bv}  
R. Brustein and G. Veneziano, \PLB 329 (1994) 429. 
\bibitem{kmo} N. Kaloper, R. Madden and K.A. Olive, \NPB 452 (1995) 677. 

\bibitem{BM} R. Brustein and R. Madden, hep-th/9702043.

\bibitem{BM2} R. Brustein and R. Madden, hep-th/9708046.

\bibitem{gmv} M. Gasperini, M. Maggiore and G. Veneziano, \NPB 494 (1997) 315. 


\bibitem{he} S. W. Hawking and G. F. R. Ellis, The large scale structure of space-time, 
Cambridge University Press, Cambridge, England, 1973

 

\end{thebibliography}
\end{document}